

\message
{JNL.TEX version 0.95 as of 5/13/90.  Using CM fonts.}

\catcode`@=11
\expandafter\ifx\csname inp@t\endcsname\relax\let\inp@t=\input
\def\input#1 {\expandafter\ifx\csname #1IsLoaded\endcsname\relax
\inp@t#1%
\expandafter\def\csname #1IsLoaded\endcsname{(#1 was previously loaded)}
\else\message{\csname #1IsLoaded\endcsname}\fi}\fi
\catcode`@=12

\font\twelverm=cmr12			\font\twelvei=cmmi12
\font\twelvesy=cmsy10 scaled 1200	\font\twelveex=cmex10 scaled 1200
\font\twelvebf=cmbx12			\font\twelvesl=cmsl12
\font\twelvett=cmtt12			\font\twelveit=cmti12
\font\twelvesc=cmcsc10 scaled 1200	\font\twelvesf=cmss12
\font\twelvemib=ambi10 scaled 1200
                     
\font\tenmib=ambi10
\font\eightmib=ambi10 scaled 800

\skewchar\twelvei='177			\skewchar\twelvesy='60
\skewchar\twelvemib='177

\newfam\mibfam

\def\twelvepoint{\normalbaselineskip=12.4pt plus 0.1pt minus 0.1pt
  \abovedisplayskip 12.4pt plus 3pt minus 9pt
  \belowdisplayskip 12.4pt plus 3pt minus 9pt
  \abovedisplayshortskip 0pt plus 3pt
  \belowdisplayshortskip 7.2pt plus 3pt minus 4pt
  \smallskipamount=3.6pt plus1.2pt minus1.2pt
  \medskipamount=7.2pt plus2.4pt minus2.4pt
  \bigskipamount=14.4pt plus4.8pt minus4.8pt
  \def\rm{\fam0\twelverm}          \def\it{\fam\itfam\twelveit}%
  \def\sl{\fam\slfam\twelvesl}     \def\bf{\fam\bffam\twelvebf}%
  \def\mit{\fam 1}                 \def\cal{\fam 2}%
  \def\sc{\twelvesc}		   \def\tt{\twelvett}%
  \def\sf{\twelvesf}               \def\mib{\fam\mibfam\twelvemib}%
  \textfont0=\twelverm   \scriptfont0=\tenrm   \scriptscriptfont0=\sevenrm
  \textfont1=\twelvei    \scriptfont1=\teni    \scriptscriptfont1=\seveni
  \textfont2=\twelvesy   \scriptfont2=\tensy   \scriptscriptfont2=\sevensy
  \textfont3=\twelveex   \scriptfont3=\twelveex\scriptscriptfont3=\twelveex
  \textfont\itfam=\twelveit
  \textfont\slfam=\twelvesl
  \textfont\bffam=\twelvebf \scriptfont\bffam=\tenbf
                            \scriptscriptfont\bffam=\sevenbf
  \textfont\mibfam=\twelvemib \scriptfont\mibfam=\tenmib
                              \scriptscriptfont\mibfam=\eightmib
  \normalbaselines\rm}


\mathchardef\alpha="710B
\mathchardef\beta="710C
\mathchardef\gamma="710D
\mathchardef\delta="710E
\mathchardef\epsilon="710F
\mathchardef\zeta="7110
\mathchardef\eta="7111
\mathchardef\theta="7112
\mathchardef\iota="7113
\mathchardef\kappa="7114
\mathchardef\lambda="7115
\mathchardef\mu="7116
\mathchardef\nu="7117
\mathchardef\xi="7118
\mathchardef\pi="7119
\mathchardef\rho="711A
\mathchardef\sigma="711B
\mathchardef\tau="711C
\mathchardef\phi="711E
\mathchardef\chi="711F
\mathchardef\psi="7120
\mathchardef\omega="7121
\mathchardef\varepsilon="7122
\mathchardef\vartheta="7123
\mathchardef\varpi="7124
\mathchardef\varrho="7125
\mathchardef\varsigma="7126
\mathchardef\varphi="7127


\def\beginlinemode{\endmode
  \begingroup\parskip=0pt \obeylines\def\\{\par}\def\endmode{\par\endgroup}}
\def\beginparmode{\endmode
  \begingroup \def\endmode{\par\endgroup}}
\let\endmode=\par
{\obeylines\gdef\
{}}
\def\singlespace{\baselineskip=\normalbaselineskip}

\def\oneandahalfspace{\baselineskip=\normalbaselineskip
  \multiply\baselineskip by 3 \divide\baselineskip by 2}
\def\doublespace{\baselineskip=\normalbaselineskip \multiply\baselineskip by 2}

\newcount\firstpageno
\firstpageno=2
\footline={\ifnum\pageno<\firstpageno{\hfil}\else{\hfil\twelverm\folio\hfil}\fi}
\def\toppageno{\global\footline={\hfil}\global\headline
  ={\ifnum\pageno<\firstpageno{\hfil}\else{\hfil\twelverm\folio\hfil}\fi}}
\let\rawfootnote=\footnote		
\def\footnote#1#2{{\rm\singlespace\parindent=0pt\parskip=0pt
  \rawfootnote{#1}{#2\hfill\vrule height 0pt depth 6pt width 0pt}}}
\def\raggedcenter{\leftskip=4em plus 12em \rightskip=\leftskip
  \parindent=0pt \parfillskip=0pt \spaceskip=.3333em \xspaceskip=.5em
  \pretolerance=9999 \tolerance=9999
  \hyphenpenalty=9999 \exhyphenpenalty=9999 }
\def\dateline{\rightline{\ifcase\month\or
  January\or February\or March\or April\or May\or June\or
  July\or August\or September\or October\or November\or December\fi
  \space\number\year}}
\def\received{\vskip 3pt plus 0.2fill
 \centerline{\sl (Received\space\ifcase\month\or
  January\or February\or March\or April\or May\or June\or
  July\or August\or September\or October\or November\or December\fi
  \qquad, \number\year)}}


\hsize=6.5truein
\hoffset=0pt
\vsize=8.9truein
\voffset=0pt
\parskip=\medskipamount
\def\\{\cr}
\twelvepoint		
\doublespace		
\overfullrule=0pt	




\def\title			
  {\null\vskip 3pt plus 0.2fill
   \beginlinemode \doublespace \raggedcenter \bf}

\def\author			
  {\vskip 3pt plus 0.2fill \beginlinemode
   \singlespace \raggedcenter\sc}

\def\affil			
  {\vskip 3pt plus 0.1fill \beginlinemode
   \oneandahalfspace \raggedcenter \sl}

\def\abstract			
  {\vskip 3pt plus 0.3fill \beginparmode
   \oneandahalfspace ABSTRACT: }

\def\endtitlepage		
  {\endpage			
   \body}

\def\body			
  {\beginparmode}		

\def\head#1{			
  \goodbreak\vskip 0.5truein	
  {\immediate\write16{#1}
   \raggedcenter \uppercase{#1}\par}
   \nobreak\vskip 0.25truein\nobreak}

\def\beginitems{
\par\medskip\bgroup\def\i##1 {\item{##1}}\def\ii##1 {\itemitem{##1}}
\leftskip=36pt\parskip=0pt}
\def\enditems{\par\egroup}

\def\beneathrel#1\under#2{\mathrel{\mathop{#2}\limits_{#1}}}


\def\references			
  {\head{References}		
   \beginparmode
   \frenchspacing \parindent=0pt \leftskip=1truecm
   \parskip=8pt plus 3pt \everypar{\hangindent=\parindent}}



\gdef\journal#1, #2, #3, 1#4#5#6{		
    {\sl #1~}{\bf #2}, #3 (1#4#5#6)}		

\def\endreferences{\body}

\def\figurecaptions		
  {\endpage
   \beginparmode
   \head{Figure Captions}
}

\def\endpage			
  {\vfill\eject}

\def\endpaper			
  {\endmode\vfill\supereject}


\def\heading				
  {\vskip 0.5truein plus 0.1truein	
   \beginparmode \def\\{\par} \parskip=0pt \singlespace \raggedcenter}

\def\subheading				
  {\vskip 0.25truein plus 0.1truein	
   \beginlinemode \singlespace \parskip=0pt \def\\{\par}\raggedcenter}

\def\tag#1$${\eqno(#1)$$}

\def\align#1$${\eqalign{#1}$$}

\def\aligntag#1$${\gdef\tag##1\\{&(##1)\cr}\eqalignno{#1\\}$$
  \gdef\tag##1$${\eqno(##1)$$}}

\def\endaligntag{}

\def\overset #1\to#2{{\mathop{#2}\limits^{#1}}}
\def\underset#1\to#2{{\let\next=#1\mathpalette\undersetpalette#2}}
\def\undersetpalette#1#2{\vtop{\baselineskip0pt
\ialign{$\mathsurround=0pt #1\hfil##\hfil$\crcr#2\crcr\next\crcr}}}


\def\ref#1{Ref.~#1}			
\def\[#1]{[\cite{#1}]}
\def\cite#1{{#1}}
\def\(#1){(\call{#1})}
\def\call#1{{#1}}
\def\taghead#1{}
\def\frac#1#2{{#1 \over #2}}

\def\12{{1\over2}}

\def\ie{{\it i.e.,\ }}

\def\sla{\raise.15ex\hbox{$/$}\kern-.57em}
\def\leaderfill{\leaders\hbox to 1em{\hss.\hss}\hfill}
\def\twiddle{\lower.9ex\rlap{$\kern-.1em\scriptstyle\sim$}}
\def\bigtwiddle{\lower1.ex\rlap{$\sim$}}
\def\gtwid{\mathrel{\raise.3ex\hbox{$>$\kern-.75em\lower1ex\hbox{$\sim$}}}}
\def\ltwid{\mathrel{\raise.3ex\hbox{$<$\kern-.75em\lower1ex\hbox{$\sim$}}}}
\def\square{\kern1pt\vbox{\hrule height 1.2pt\hbox{\vrule width 1.2pt\hskip 3pt
   \vbox{\vskip 6pt}\hskip 3pt\vrule width 0.6pt}\hrule height 0.6pt}\kern1pt}
\def\tdot#1{\mathord{\mathop{#1}\limits^{\kern2pt\ldots}}}

\def\pmb#1{\setbox0=\hbox{#1}%
  \kern-.025em\copy0\kern-\wd0
  \kern  .05em\copy0\kern-\wd0
  \kern-.025em\raise.0433em\box0 }

\catcode`@=11
\newcount\tagnumber\tagnumber=0

\immediate\newwrite\eqnfile
\newif\if@qnfile\@qnfilefalse
\def\write@qn#1{}
\def\writenew@qn#1{}
\def\w@rnwrite#1{\write@qn{#1}\message{#1}}
\def\@rrwrite#1{\write@qn{#1}\errmessage{#1}}

\def\taghead#1{\gdef\t@ghead{#1}\global\tagnumber=0}
\def\t@ghead{}

\expandafter\def\csname @qnnum-3\endcsname
  {{\t@ghead\advance\tagnumber by -3\relax\number\tagnumber}}
\expandafter\def\csname @qnnum-2\endcsname
  {{\t@ghead\advance\tagnumber by -2\relax\number\tagnumber}}
\expandafter\def\csname @qnnum-1\endcsname
  {{\t@ghead\advance\tagnumber by -1\relax\number\tagnumber}}
\expandafter\def\csname @qnnum0\endcsname
  {\t@ghead\number\tagnumber}
\expandafter\def\csname @qnnum+1\endcsname
  {{\t@ghead\advance\tagnumber by 1\relax\number\tagnumber}}
\expandafter\def\csname @qnnum+2\endcsname
  {{\t@ghead\advance\tagnumber by 2\relax\number\tagnumber}}
\expandafter\def\csname @qnnum+3\endcsname
  {{\t@ghead\advance\tagnumber by 3\relax\number\tagnumber}}

\def\equationfile{%
  \@qnfiletrue\immediate\openout\eqnfile=\jobname.eqn%
  \def\write@qn##1{\if@qnfile\immediate\write\eqnfile{##1}\fi}
  \def\writenew@qn##1{\if@qnfile\immediate\write\eqnfile
    {\noexpand\tag{##1} = (\t@ghead\number\tagnumber)}\fi}
}

\def\callall#1{\xdef#1##1{#1{\noexpand\call{##1}}}}
\def\call#1{\each@rg\callr@nge{#1}}

\def\each@rg#1#2{{\let\thecsname=#1\expandafter\first@rg#2,\end,}}
\def\first@rg#1,{\thecsname{#1}\apply@rg}
\def\apply@rg#1,{\ifx\end#1\let\next=\relax%
\else,\thecsname{#1}\let\next=\apply@rg\fi\next}

\def\callr@nge#1{\calldor@nge#1-\end-}
\def\callr@ngeat#1\end-{#1}
\def\calldor@nge#1-#2-{\ifx\end#2\@qneatspace#1 %
  \else\calll@@p{#1}{#2}\callr@ngeat\fi}
\def\calll@@p#1#2{\ifnum#1>#2{\@rrwrite{Equation range #1-#2\space is bad.}
\errhelp{If you call a series of equations by the notation M-N, then M and
N must be integers, and N must be greater than or equal to M.}}\else%
 {\count0=#1\count1=#2\advance\count1
by1\relax\expandafter\@qncall\the\count0,%
  \loop\advance\count0 by1\relax%
    \ifnum\count0<\count1,\expandafter\@qncall\the\count0,%
  \repeat}\fi}

\def\@qneatspace#1#2 {\@qncall#1#2,}
\def\@qncall#1,{\ifunc@lled{#1}{\def\next{#1}\ifx\next\empty\else
  \w@rnwrite{Equation number \noexpand\(>>#1<<) has not been defined yet.}
  >>#1<<\fi}\else\csname @qnnum#1\endcsname\fi}

\let\eqnono=\eqno
\def\eqno(#1){\tag#1}
\def\tag#1$${\eqnono(\displayt@g#1 )$$}

\def\aligntag#1\endaligntag
  $${\gdef\tag##1\\{&(##1 )\cr}\eqalignno{#1\\}$$
  \gdef\tag##1$${\eqnono(\displayt@g##1 )$$}}

\def\eqalignno#1{\displ@y \tabskip\centering
  \halign to\displaywidth{\hfil$\displaystyle{##}$\tabskip\z@skip
    &$\displaystyle{{}##}$\hfil\tabskip\centering
    &\llap{$\displayt@gpar##$}\tabskip\z@skip\crcr
    #1\crcr}}

\def\displayt@gpar(#1){(\displayt@g#1 )}

\def\displayt@g#1 {\rm\ifunc@lled{#1}\global\advance\tagnumber by1
        {\def\next{#1}\ifx\next\empty\else\expandafter
        \xdef\csname @qnnum#1\endcsname{\t@ghead\number\tagnumber}\fi}%
  \writenew@qn{#1}\t@ghead\number\tagnumber\else
        {\edef\next{\t@ghead\number\tagnumber}%
        \expandafter\ifx\csname @qnnum#1\endcsname\next\else
        \w@rnwrite{Equation \noexpand\tag{#1} is a duplicate number.}\fi}%
  \csname @qnnum#1\endcsname\fi}

\def\ifunc@lled#1{\expandafter\ifx\csname @qnnum#1\endcsname\relax}

\let\@qnend=\end\gdef\end{\if@qnfile
\immediate\write16{Equation numbers written on []\jobname.EQN.}\fi\@qnend}

\catcode`@=12


\title
Dynamics of Earthquake Faults
\author
J.M. Carlson,$^{*{\dag}}$ J.S. Langer,$^{\dag}$ and B.E. Shaw$^{{\dag}\circ}$

\abstract
We present an overview of our ongoing studies of the
rich dynamical behavior
of the uniform, deterministic Burridge--Knopoff model  of an
earthquake fault.
We discuss the behavior of the model in the context of
current questions in seismology.
Some of the topics considered include:
(1) basic properties of the model, such as the magnitude vs.~frequency
distribution and the distinction between small and large events;
(2) dynamics of individual events, including dynamical selection
of rupture propagation speeds;
(3) generalizations of the model to more realistic, higher dimensional
models;
(4) studies of predictability, in which artificial
catalogs generated by the model are used to test and determine
the limitations of pattern recognition algorithms used in seismology.

\vskip 1.truein
\noindent
$^*$ Department of Physics, University of California, Santa Barbara, CA 93106

\noindent
$^{\dag}$Institute for Theoretical Physics, University of California,
Santa Barbara, CA 93106

\noindent
$^{\circ}$ Lamont-Doherty Earth Observatory, Columbia University, Palisades, NY
10964

\endtitlepage

\head{I. Introduction: The dynamical richness of seismic phenomena}

Two trends that have characterized much of modern theoretical physics
are an increased capability for dealing with complex systems and an
increased interest in topics that traditionally have been the
property of other disciplines.  Investigations that fall into the
latter category can be specially rewarding, but it is necessary to
``take stock" every so often to make sure that one is
really working along lines that are meaningful. The present article
is a ``stock-taking" with regard to our recent investigations of the
dynamics of a simple model of an earthquake
fault.

The earthquake problem certainly satisfies our criteria for
trendiness both in complexity and interdisciplinarity. A wide variety
of points of view have been taken in recent years by geologists,
seismologists, mechanical engineers,
mathematicians and
physicists.
Ours has been a comparatively narrow
one. We have focused on what we believe to be the simplest nontrivial
model of a single segment of an earthquake fault and have
examined its dynamic properties. As we  describe below, our
results have been interesting enough to encourage us to extend our
investigations to more practical applications of the model. Before
outlining these
results, however, we will provide a broader context for them by
devoting a few paragraphs to some general aspects of the earthquake
problem. For a more complete discussion see Scholz [1990].

One  spectacular feature of earthquakes is the enormous
range of scales over which they occur.
The distribution of sizes of seismic events is observed to be a
power law over more than ten orders of magnitude (though
the exponent may vary somewhat across this range).
This power law distribution of event sizes, known as the Gutenberg--Richter
law [Gutenberg and Richter, 1954],
is one of the most fundamental observations in seismology.
Its explanation  remains an outstanding question
in the earth sciences. Two possible origins have  been
the topic of much debate recently.
One point of view is that
geometric and material irregularities in the earth dominate the behavior,
and it is therefore  \lq\lq quenched disorder"
which
underlies the answer.
An alternative point of view is one in which
complexity arises primarily as a consequence of
dynamical instabilities. In this case, nonuniformities evolve
as the system evolves, with no {\it a priori}
fixed inhomogeneities or stochastic forcing.
Whether the
Earth is operating in one regime or another or in some combination
remains an open and fundamental question.

Typically, earthquakes occur in the upper tens of kilometers
of the Earth's crust.
They arise as a consequence of frictional instabilities
which cause stress, accumulated by large scale plate motions over periods of
hundreds of years, to
be relieved in sudden stick-slip events.
In other cases,  frictional properties may allow
stable sliding, so that strain is relieved smoothly and aseismically.
A different class of events, known as \lq\lq deep earthquakes"
occurs along subducting plates, at depths of hundreds
of kilometers. The  basic physical processes which are
responsible for deep earthquakes are
not well understood, but these events are suspected to be
caused by  the plates undergoing phase changes at the
high temperatures and pressures associated with large depths.

The motions of tectonic plates are driven by large-scale convective
flows in the lower mantle. These flows take place on length scales of
thousands of kilometers with turnover times of hundreds of millions
of years; thus, characteristic speeds are of the order of centimeters
per year. Understanding these inner motions of the Earth is yet
another important object of research, carried out these days
primarily by geophysicists and fluid dynamicists [Glatzmaier, {\it et al.,}
1990].
These inner motions
may be intrinsically chaotic; for example, material from the Earth's
core may be brought to the surface by intermittent plumes that rise
through the mantle. Nevertheless, the large scale convection is so
slow and
powerful that it may be thought of as providing a constant external
driving force for seismological purposes.

Of more direct seismological interest is the manner in which the
inner flows couple to the brittle outer layers, thereby driving
relative motions of the plates and producing the intricate patterns
of cracks that we know as earthquake faults. One of the main themes
of modern seismology is the use of basic principles of fracture
mechanics for understanding the way in which stresses applied to the
crust have produced the various kinds of faults that are known to
exist (see, e.g., Scholz [1990]). A recent trend has been to
ascribe fractal properties to the
complex arrays of fault segments that occur in the regions where
major plates come into contact with each other
[Kagan, 1982; Barriere, {\it et al.}, 1991; Sahimi, {\it et al.,} 1993;
Knopoff, 1993].
Outstanding research
questions are why these patterns have the geometric  properties that
are observed, and what may be the implications of these geometries
for predicting the frequencies and sizes of seismic events. It may
be, for example, that each segment of a complex pattern of faults
undergoes its own characteristic cycle of earthquakes with only
relatively weak coupling to its neighbors. If so, then
the statistical distribution of earthquake magnitudes
will be determined primarily by the statistics of fault segments--
the disorder is quenched into the system-- and the most
important concepts in theoretical seismology will be geometric in nature.

A unifying hypothesis that may encompass both the geometric point of
view  and the dynamic picture to be emphasized here is
known as ``self-organized criticality" [Bak, {\it et al.,} 1987;
Chen, {\it et al.,} 1991].
The idea is that many systems
in nature, the Earth's crust for example, are driven by external
forces in such a way that they are always at or near a threshold of
instability.
Tectonic plates retain their integrity or remain locked
to one another until the stresses that are imposed upon them are
partially relieved by events such as fault formation or earthquakes,
but the stresses begin to rise toward threshold again as soon as an
event is over. It seems possible that
systems that operate persistently near a threshold of instability are
in some way like thermodynamic systems near critical points. If, as
in critical phenomena, some length or time scale is diverging near
threshold, then fluctuations may occur over anomalously broad ranges
of size or duration. So far as we know, there does not yet exist a
systematic mathematical definition of a state of self-organized
criticality. The hypothesis remains an intriguing
conjecture that has served to stimulate a variety of theoretical
investigations including our own. As we will see, the earthquake
model that we have studied does exhibit some features of a critical
system, but there are other respects in which it differs markedly
from the hypothetical ideal.

The principal result of our studies is that a spatially uniform
block-and-spring model --- the one-dimensional ``Burridge-Knopoff
model'' [Burridge and Knopoff, 1967] without built-in irregularities or
external noise --- is a
deterministically chaotic dynamical system whose behavior is similar
in some important ways to the behavior of real earthquake
faults.   In
particular, the model exhibits a broad spectrum of small to
moderately large, spatially localized, earthquake-like events which
move stress from one location to another but do very little to
relieve stress in the system as a whole
[Carlson and Langer, 1989a, 1989b; Carlson, {\it et al.}, 1991].
It also exhibits spatially
extended ``great events'' in which strong slipping pulses arise at
``epicenters'' and propagate along the fault, thereby unloading the
stress on large segments of the system.  The sizes of the
smaller events obey a power-law distribution much like the
Gutenberg-Richter law.
The great events recur irregularly on roughly
a loading period --- the time necessary for the tectonic forces to
build stress up to the breaking point --- and their frequency as a
function of magnitude is not generally a simple extension of the
power-law distribution for the smaller events [Carlson, 1991].
This rich pattern of
apparently realistic behavior emerges from a model with essentially only one
adjustable, \ie material-dependent parameter, and with no externally
imposed spatial or temporal structure.

We have used the term ``apparently'' here because this is, at best, a
model of a single isolated fault or fault segment.  The intrinsic
dynamic properties of single faults are not known with precision, and
many aspects of the subject remain controversial.  The catalogs of
seismic events on which the Gutenberg-Richter law is based almost
always pertain to large regions containing many faults; and the
statistics of large characteristic events on single faults are
necessarily very poor because the recurrence times for such events
are usually centuries or longer [Nishenko and Buland, 1987].
The manner in which slip propagates
in this model [Langer and Tang, 1991; Langer, 1992; Myers and Langer 1993a;
Langer, 1993;
Langer and Nakanishi, 1993] looks very much like the
conjectured ``Heaton pulse'' [Heaton, 1990].
However, much work remains to be done before we will know whether this
pulse is actually the normal mechanism for slip in large earthquakes
and, if so, whether the pulse that occurs in the Burridge--Knopoff
model has anything in
common with the real one.

Thus, there are
large uncertainties in our understanding of seismic
phenomena in the Earth, and the uniform
Burridge-Knopoff model is far from being an accurate representation
of even the simplest faults. (It does not even contain a mechanism for
aftershocks.) What, then, can we say about
meaningful accomplishments and future prospects for research along
these lines?  In large part, this article is devoted to answering
that question.  As in any theoretical project, much of the answer is
that the insight gained from modelistic studies can be helpful in
interpreting real data and in developing more realistic models.
After a brief introduction to the simplest version of this model in
Section II, we describe several such interpretations and extensions
that we believe are particularly promising: the distinction between
localized and delocalized events, pulse propagation, and
some  higher dimensional models which include either
patterns of slip on the fault plane or elastic deformations normal to
that plane.  These topics are discussed in Sections III - V.

Finally, we conclude in
Section VI with an introduction to what is --- for physicists, at
least --- an unconventional use of physical models.  The idea is to
use this model as a tool for developing objective techniques for
earthquake prediction.  As mentioned above, the Burridge-Knopoff
model is highly idealized, but it has two specially important
features: it is intrinsically and deterministically chaotic -- and
therefore technically ``unpredictable'';  and it produces a clear
distinction between frequent small events and the rare large ones
about which we would like to be warned in advance.  Thus, this model
and its more realistic extensions ought to provide practical,
quantitative tests for the validity of prediction schemes.

\head{II. The Model: Complex behavior from a simple dynamics}

	We start by considering the simple model of a
lateral earthquake fault illustrated schematically in
Fig.~1. Here, two adjoining elastic plates are being
pushed together and forced to move in opposite directions
along their line of contact.  The corresponding
compressive and shear stresses are applied at some
distance from this ``fault'' line.  The plates are
``stuck'' to each other at the fault but, whenever the
threshold for sticking friction is exceeded by the local
shear stress, the plates move in an earthquake-like
slipping event.

	This model contains three essential ingredients:

(1) A mechanism for loading the system, i.e. for applying
the shear stress.

(2) Mechanisms for storing the elastic energy associated
with both the compressive and shear stresses.

(3) Stick-slip friction acting between the plates along
the fault line.  For our purposes, it is essential that
this be a velocity-weakening friction.  That is, once
slipping begins, the frictional force decreases with
increasing slipping speed. The resulting dynamic
instability is responsible for almost all of the
interesting properties of this class of models.

We also insist that this be a fully deterministic,
dynamical model.  The benefit of studying a dynamical
system as opposed to, say, a cellular automaton is that
we can more easily identify characteristic length and
time scales with the corresponding parameters in the
laboratory or the  Earth.  Moreover, our physical
intuition leads us to believe that deterministic inertial
dynamics is an essential ingredient of a theory of
earthquakes.

	All of these ingredients are contained in the
uniform, one-dimensional Burridge-Knopoff model, which we
can define by the following partial differential
equation:
$$
\ddot U = \xi^2 {\partial^2 U\over\partial x^2} - U -
\phi(\dot U) + \nu\,t.\eqno(eom)
$$
The original Burridge-Knopoff model is the
one-dimensional chain of blocks and springs illustrated
in Fig.~2; the equations of motion for that system are
identical to the elementary finite-difference
approximation for \(eom).  Equation \(eom), or its discretized
version, is Newton's law of motion in an almost
dimensionless form.  The function $U(x,t)$ is the time-dependent
displacement of the material at position $x$
along the fault.  The inertial term is on the left-hand
side.  The first two terms on the right are the (linear)
compressional and shear forces, respectively, and
$\phi(\dot U)$ is the velocity-dependent friction.
Finally, $\nu\,t$ is the driving force; that is, $\nu$ is
the dimensionless loading rate.  Note that, in contrast
to the original work by Burridge and Knopoff and many
subsequent investigators, our version of the model is
completely uniform.  All of the physical elements ---
masses, springs, friction forces --- are identical to one
another; and we shall introduce no external noise or
other stochastic ingredients.

	We have scaled Eq.\(eom) in such a way that the
largest possible earthquake, in which the system slips
forward uniformly through one half period of the simple
harmonic motion determined by the shear force, has a
duration $\Delta t = \pi$ and a corresponding slip
distance $\Delta U = 2$.  In other words, our time $t$ is
measured in units of a characteristic slip time
(sometimes called the ``rise time'') which, in the Earth,  is
of order seconds. Our displacements $U$ are measured
in units of a characteristic slip distance, which usually
is of order meters.

	The dimensionless loading rate $\nu$ is measured in
units of the characteristic slipping speed.  Because the
latter is of order meters per second, and geological
loading rates are of order centimeters per year, $\nu$ is
of order $10^{-8}$ or less.  An equivalent way of saying
this is that geological loading periods --- the times
required for the tectonic forces to build back up to the
slipping threshold after a very large event --- are
generally of the order centuries in real time, and are of
the order $\nu^{-1}\approx 10^8$ in our dimensionless
units.

	In order to avoid confusion between displacement $U$
and position $x$ (which both have the dimensions of
length but which scale quite differently) we have chosen
to leave $x$ in dimensional units rather than further
simplifying Eq.\(eom) by writing $x$ in units of $\xi$.
The length $\xi$ is the distance traveled by a sound
wave in a slipping time, and thus is of order ten
kilometers.  This length necessarily is the same as the
thickness of the seismogenic layer of the Earth's crust;
it is the only length available for setting the
frequency of the slowest harmonic mode in the system,
that is, for determining the coefficient (unity in our notation) of
$-U$ in \(eom). Because time is
dimensionless, we will also speak of $\xi$ as a velocity.
The one remaining length scale in the
problem is the block spacing $a$, which is shown in
Fig.~2, but which disappears in the continuum limit
implicit in \(eom).  We shall see that $a$ plays a
crucial role as a short-distance cutoff; that is, \(eom) is not by itself a
mathematically well posed differential equation.

	The only nonlinearity in Eq.\(eom) is contained in
the slip-stick friction law  $\phi(\dot U)$, which we
show schematically in Fig.~3.  An elementary mass point
along the fault --- that is, a ``block'' in Fig.~2 ---
remains stuck at $\dot U = 0$ until the force on it
exceeds some threshold, which we have scaled to unity.
When a block starts to slide, the frictional resistance
to its motion decreases at a rate $2\alpha$. The
quantity $\alpha$ is (very nearly) the only dimensionless group of parameters
that contains information about the constitutive properties of the material; it
is the ratio of
the characteristic slipping speed (meters per second) to
a speed that characterizes the velocity-weakening
property of the friction law.  Thus, if $\alpha$ is
large, the fault quickly becomes ``slippery'' once motion
begins, and this motion is strongly unstable.
Conversely, if $\alpha$ is small, the fault remains
relatively ``sticky'' during motion, and the instability
is weak.  It is important to note that, because the
elastic forces in Eq.\(eom) are all linear, we can reset
the zero of $\phi(\dot U)$ simply by shifting the zero of
$U$; thus, without loss of generality, we have let
$\phi(\dot U)$ vanish for large $\dot U$, and we also
have let the multi-valued quantity $\phi(0)$ be as large
and negative as is necessary to prevent back-sliding.
That is, we constrain the motion so that $\dot U \ge 0$.

	For the numerical calculations reported here, we
have chosen
$$
\phi(\dot U)=\cases{(-\infty,0], & $\dot U=0$\cr
                 {(1-\sigma)\over 1+{(2\alpha\dot
U/(1-\sigma))}}, & $\dot U>0$.\cr}\eqno(friction)
$$
The ``onset parameter'' $\sigma$ is the acceleration of a
block at the instant when slipping begins. We have
introduced $\sigma$ primarily as a technical device that
allows us to study the limit $\nu\to 0$. In the absence
of $\sigma$, events begin indefinitely slowly in that
limit, which obviously is inconvenient for numerical
purposes.
Further, it seems  improbable that the order of magnitude
of the very  small
driving rate $\nu$  should influence the
dynamics of  events on a completely different scale.
In fact, if there is a separate nucleation process
leading up to an event, then
a finite $\sigma$, which provides an  initial \lq\lq kick,"
may be a realistic effect.
With nonzero $\sigma$ and vanishingly small
$\nu$ we need not carry out explicit
time integrations between slipping events, and thus we
are able both to use a geologically realistic loading
rate and study the system for very large numbers of
loading cycles.  Accordingly, we have used small values
of $\sigma$, usually $10^{-2}$, for generating large
artificial earthquake catalogs.

	In summary, the uniform Burridge-Knopoff model is
fully determined by only the following dimensionless,
system-dependent parameters.  Principally, there is the
friction parameter $\alpha$ which
determines all of the most important features of
earthquakes, especially the qualitative features of their
frequency distribution as a function of magnitude.  In
addition, there is the ratio $\xi/a$
which diverges in the continuum
limit, and determines
the range of sizes
over which different behaviors are seen.
As might be expected, $\xi/a$ determines the
small-magnitude cutoff of the frequency distribution.  Interestingly, it
also determines the upper cutoff for great events in very
large systems and, in addition, the
propagation speeds and widths of rupture pulses in such
events.
The loading rate $\nu$ and the onset parameter
$\sigma$ may both be thought of as being arbitrarily
small.  (They do, however, play roles in distinguishing
localized from delocalized events.) The only other
parameter that plays any role is the size of the system,
that is, the number of ``blocks''.  There are some
situations in which it is useful to consider small
systems; for example, it may be interesting to simulate
the behavior of short fault segments.  For present
purposes, however, we shall consider only systems that are
indefinitely large; then, because there is a finite cutoff $\mu^*$ for the
largest events, in large enough systems the system size is
irrelevant.

\head{III.  Properties of the uniform Burridge--Knopoff
model: Small and Large Events}

	In our earliest investigations of the uniform
Burridge-Knopoff model, we have taken advantage of
the ease with which \(eom) can be integrated
numerically to generate long and accurate artificial
catalogs of seismic events.  Using only a modest
workstation, we have been able to obtain the
geological equivalent of tens of thousands of  years of data
which, unlike real seismological data, are free of
observational errors or uncertainties.  Our initial
goal in this effort was simply to find out whether so
elementary a model might behave in a way that is at
all similar to the behavior of a real earthquake
fault.  We shall describe a more ambitious and
speculative use of these artificial catalogs in
Section VI.

	We generally have carried out our numerical
solutions of \(eom) by choosing some small inhomogeneity
in the initial displacements
and then allowing the system to
evolve until it reaches a statistically steady state.
A typical evolution of the system, beginning after
the initial transient has passed, is illustrated in
Fig.4.  Here, we plot a sequence of ``stuck''
configurations $U(x,t)$ as functions of position $x$.
A new curve is drawn after each event, which begins
when the force on some block exceeds the static
friction threshold and concludes when the entire
system comes to rest.  Clearly, the motion shown here
is very complex.  Indeed, it is technically chaotic
in the sense that nearby configurations move apart, on
average, exponentially fast.
Events occur over a wide range of sizes and, while
the eye can discern certain patterns,  the system is
not behaving in a periodic or quasiperiodic manner.

	The most striking feature of Fig. 4 is
the
distinction between small and large events. The
smaller events are by far the most numerous.  They
tend to occur in clusters and to fill in local minima
in the displacement curve $U(x,t)$ where many blocks
are close to their slipping thresholds. The large
events are much less frequent but dominate
the net motion of the system; they cover
most of the area in Fig. 4.

	In analogy to the seismological convention, we
define the seismic moment $M$ to be the integrated
slip, that is, the area swept out by an event in
Fig.4.  We further define the ``magnitude''
of an
event\footnote{$^1$}{The
usual Richter magnitude is measured on a log$_{10}$ scale, and
for historical reasons
is defined in terms of the amplitude of radiated waves at a
period of about 20 seconds. This period  corresponds to the
time it takes a sound wave
to travel a distance of order the crust depth.
The magnitude $\mu$ which we use is most closely related to a quantity
referred to as the moment magnitude, which is based on
estimates of the total slip.  The wide range of
sizes of seismic events, and the necessarily remote means by which
events are detected, has led to the development of numerous
magnitude scales. The seismic moment is the preferred measure
because it is the conserved quantity associated with
the net relative plate motion.}
to be $\mu=\ln M$,
and
we denote the differential magnitude vs.~frequency
distribution by $R(\mu)$. A typical frequency
distribution $R(\mu)$ is shown in Fig.5 for a
sequence of events like that shown in Fig.4, but for
a much larger system (over 8000 blocks),  a much larger
number of loading cycles (over 100), and slightly different
values of the parameters.  Here we see that the
sharp distinction between small and large events that
is visible in Fig.4 is also manifest in the
statistical properties of the system.  The small
events, with magnitudes between some lower bound
$\mu_1$ determined by the block size and a crossover
value $\tilde\mu$, are distributed according to our
analog of the classic Gutenberg-Richter law:
$$
R(\mu)=Ae^{-b\mu}\eqno(GR)
$$
with $b$ very nearly equal to unity for large
enough $\alpha$. (For ``stickier'' faults with $\alpha$
less than about 2,  smaller values of $b$ are obtained [Carlson and Langer,
1989b;
Vasconcelos, {\it et al.,} 1992].)  The large
events, with magnitudes greater than  $\tilde\mu$,
occur at a rate that is
higher that the extrapolated ``$b=1$'' law, and their
distribution cuts off sharply at a magnitude $\mu^*$
which, for large enough systems, is independent of
the system size.

	All of the above features of Fig.5 are
qualitatively similar to behavior that is often ---
but not always --- observed seismologically.
The
similarity pertains only to data taken for a single
fault system and not to catalogs that combine data
from many different faults with qualitatively
different characteristics (see, e.g., Scholz [1990]).  The interpretation of
earthquake catalogs remains somewhat controversial,
and we shall not try to discuss in any detail the
current state of those debates.  In brief, it appears
that smaller events on real faults, with Richter
magnitudes roughly in the range 1 - 6, are in some
sense self-similar and obey some scaling
distribution, but probably with $b$-values (as
defined here) appreciably smaller than unity [Pacheco, {\it et al.,} 1992].
The
overabundance  of large events for single faults is
difficult to observe directly because repeat times are
of the order of hundreds of years.  However, the
distribution of smaller events, even if extrapolated
to Richter magnitudes 8 or 9, appears to account for
only a small fraction of the total slip on individual
faults.  Thus there is indirect geological evidence
that most of the displacement and energy release
occurs during very large ``characteristic''
earthquakes whose recurrence frequency lies well
above the extrapolated scaling distribution
[Schwartz and Coppersmith, 1984;
Davison and Scholz, 1985; Wesnousky, 1993].

	The distinction between large and small --- or
delocalized and localized --- events is probably the
most significant new physical insight to emerge from
this study.  We can understand what is happening
roughly as follows.  Consider a zone along the fault
of width $\Delta x$ in which all of the blocks are at
or very near their slipping thresholds; that is,
in Eq.~\(eom) $\xi^2\partial^2 U/\partial x^2 - U = 1$.
When we linearize \(eom) about a configuration of
this kind, we find that an initially sharp slipping
pulse propagates at speed $\xi$ and grows like
$e^{\alpha t}$ as it picks up stored elastic energy
within the zone.  If triggered by the natural
evolution of the system, the initial amplitude of
such a pulse would be proportional to the onset
parameter $\sigma$. The delocalization criterion is
simply the condition that, when this growing pulse
reaches the edge of the zone where the blocks are
further away from their slipping thresholds, its
strength has become sufficient to dislodge those
blocks and thus sustain its motion.  Because the
pulse propagates for a time of order $\Delta x/\xi$
within the zone, the width of the smallest slipping
zone that can nucleate a delocalized event must be
proportional to $(\xi/\alpha){\rm ln}(1/\sigma)$.

	A more careful analysis [Carlson and Langer,
1989b; Carlson, {\it et al.,} 1991]
yields the delocalization length:\footnote{$^2$}
{We have confirmed
numerically that  the delocalization length
$\tilde\xi$ is proportional to $\xi$. However, the $\alpha$-dependence
appears to be  less strong than (4) would suggest.
Tests of the weak logarithmic dependence on the
additional parameters have not been carried out due to the
limited ranges over which
these parameters can be adjusted  while maintaining
good statistics.}
$$
\tilde\xi \approx
\left(2\xi\over\alpha\right) \ln\left(4\xi^2\over\sigma a^2\right)
\eqno(xitilde)
$$
Integrating the displacement associated with an event
of size $\tilde\xi$, we obtain the crossover
magnitude
$$
\tilde\mu\approx\ln\left(2\xi\over\alpha\right), \eqno(mutilde)
$$
which is consistent with the minimum in Fig.5.

	For realistic values of the parameters, the
crossover length $\tilde\xi$ is of order tens of
kilometers.  Because $\tilde\xi$ scales like $\xi$,
it is generally going to be of the same order of
magnitude as the thickness of the seismogenic layer
even in this one-dimensional
model where earthquakes have no structure
in the fault plane.
The length
$\tilde\xi$ can be viewed as a correlation length for
the model because it defines an upper bound on the
length scale over which small events tend to cluster.
Events which occur on length scales smaller than
$\tilde\xi$ tend to smooth the system, thereby
preparing an increasingly large triggering zone for a
coming large event. On the other hand,
because of the stick-slip instability, events which
occur on length scales larger than $\tilde\xi$ tend
to roughen the system, leading to later sequences of
smaller events. This interplay between small and
large events implies that an understanding of the
dynamics of the large events will be necessary in
order to calculate the scaling distribution for the
smaller events.  As yet, we have found no convincing
derivation of the ``$b=1$'' law.

Note  that the block
spacing $a$ appears explicitly on the right-hand side of
\(xitilde); this is our first indication that the
short-wavelength cutoff is playing an important role
in the dynamics of this system.
Interestingly, our numerical simulations indicate that
$a$ plays a role in determining
the size of
the largest event $\mu^*$ as well. (See Fig.5.)
We will see in the following section that $a$ shows up again
in the pulse propagation problem.

\head{IV. Ruptures: Numerical observations and analytical solutions}

Two unavoidable facts make direct
observation of
earthquake ruptures nearly impossible.
First, most earthquakes occur well below the surface of the Earth,
and, second, earthquakes occur suddenly, at irregular time intervals,
making it difficult
for the observer to be at the right place at the right time.
Thus, instead of direct measurements,  seismologists
measure the shaking produced at the Earth's surface at remote stations, and
try to invert the signal
to infer the motions at the source.
However, the inversion problem is intrinsically complicated.
For example,  the effects of inhomogeneities in the crust on
the attenuation of  radiated
waves
are  not fully understood,
making it difficult to separate the effects which are due to
the source from those due to  the path.
In addition, the problem of inverting an array of seismograms to deduce
the source motion is underdetermined since it involves
reconstruction of motions in two space and one time dimension using
a finite number of one-dimensional time series collected at
different   stations.
Theoretical models thus play a crucial role in the
inversion process.

In the Burridge--Knopoff model we can use our privileged position of being able
to follow the motion directly to take a closer look at
the dynamics of individual events.
A very large
delocalized event is illustrated in Fig.6. Here we show a sequence of
configurations at equally spaced time intervals during
the period in which slipping is actually taking place.
The motion begins at the point labeled ``epicenter''
and looks initially like a localized event in which
the blocks in a zone roughly of width $\tilde\xi$ slip
forward irregularly.  As the motion reaches the edges
of the zone, it becomes a pair of smooth
pulses that propagate outward at almost constant
speed.  Each nearly vertical line in Fig.6 is a
slipping region in which the blocks are jumping
forward from one stuck configuration (the bottom
curve) to another (the top). The right-moving pulse
dies out quickly because it encounters a nearby region
in which the blocks are firmly stuck. The left-moving
pulse, on the other hand, moves much farther, slowing
slightly and then accelerating again as it passes
through stuck regions, and finally coming to rest only
after relieving the stress over a large portion of the
fault.

Numerically we can follow the net motion forward
as a function of time and, taking the Fourier transform of this,
obtain the moment spectrum, a quantity that seismologists can measure.
The spectra produced by the model exhibit power laws and are
similar to spectra taken from the Earth. Interestingly, $\tilde\xi$
is again  relevant.
The model spectra for large events exhibit a bend at a
frequency proportional to the inverse time required for a pulse to
propagate a distance $\tilde\xi$
[Shaw, 1993a].

Even more powerfully, analytic calculations have been made
of the speed and shape of pulses propagating into a uniformly
stuck state. The shape of  slip pulses in the Earth remains
an open question, but one which is an important ingredient
in the inversion process.  Some
simple considerations make it clear that even in our simple
model this problem
is non-trivial; indeed, just posing the problem leads
us to explore some previously untouched but
fundamental issues in fracture dynamics [Langer, 1992, 1993; Langer and
Nakanishi, 1993].  In the brief
remarks that follow, we can provide only a very
qualitative and nonrigorous summary of our ongoing
investigations in this area.  More details can be
found in Langer and Tang [1991] and  Myers and Langer [1993a].

	Without loss of generality for these purposes,
we can set $\nu t = 0$ in \(eom) and, for simplicity,
consider a pulse propagating into a system that is
uniformly at, say, $U=-1+\epsilon$. Because the
slipping threshold occurs at $\phi=1$, $\epsilon$ is
our measure of distance from threshold, that is, it is
the degree to which the system is stuck. Our
problem is to find steady-state propagating solutions
of the form $U(x,t)=U(x-vt)$ with $U(x\gg vt)\approx -
1+\epsilon$, and to compute the speed $v$ as a
function of $\epsilon$.

	The simplest and most naive approach to this
problem is to assume that the slipping friction
$\phi(\dot U >0)$ drops quickly to zero, in which case
solutions of \(eom) have the form:
$$
U(x,t)\approx \cases{1-\epsilon &${x-vt\over
\sqrt{v^2-\xi^2}} < -{\pi\over 2}$\cr
 -(1-\epsilon)\sin \left(x-vt\over \sqrt{v^2-
\xi^2}\right) &$-{\pi\over 2}<{x-vt\over \sqrt{v^2-
\xi^2}}<{\pi\over 2}$\cr -1+\epsilon &${x-vt\over
\sqrt{v^2-\xi^2}}>+{\pi\over 2}$\cr}\eqno(pulse)
$$
Despite the fact that the slipping friction cannot be
completely negligible at the onset of motion and at
the resticking point, and despite the fact that we
have written \(pulse) without proper consideration of
the boundary conditions at these points, this result
turns out to be remarkably close to the truth. Formal
steady-state solutions of properly posed
interpretations of \(eom) can be shown to exist for a
continuous range of supersonic speeds $v>\xi$; and the
slipping motion, to a good approximation, is a free-
slipping pulse much like \(pulse) whose width is
proportional to $\sqrt{v^2-\xi^2}$. The numerical
experiments, however, imply not a family of values of
$v$ but sharp selection independent of initial
conditions.

	The crucial, physically important fact about the
velocity-selection problem is that its solution
requires the introduction of a new length scale or,
equivalently, a short-wavelength cutoff.  We have
dealt with this situation  in two different ways.  In
our earliest work on this problem, we used the fact
that the finite-difference approximation to \(eom),
that is, the uniform block and spring model of
Burridge and Knopoff, is perfectly well posed
mathematically.  For small block spacing $a$, a
sufficiently accurate finite-difference correction can
be obtained by adding the term $(\xi^2 a^2/12)
\partial^4 U/\partial x^4$ to the right-hand side of
\(eom).  In recent investigations, we have preferred
to add a viscous damping of the form $\eta
\partial^2\dot U/\partial x^2$ on grounds that such a
term has a more natural physical interpretation. This
viscous term becomes $-\eta v\partial^3U/\partial x^3$
when we look only for steady-state solutions.  The
relevant new length in this case is $\sqrt \eta$. Both
techniques amount to adding a comparatively high
derivative to the equation of motion, and thus
introducing a singular perturbation that does in
fact make the equation mathematically well defined.
For simplicity, we shall refer here only to the
finite-difference method, although it is the viscous
model that has proved to be most interesting in a
number of physical contexts.

	With the addition of the singular perturbation
and the associated new length scale, velocity
selection in this system becomes a non-standard version of a
specially interesting class of dynamic phenomena.  The
history of work in this area goes back to the classic
paper of Kolmogorov, Petrovskii, and Piscounov [1937] on
front propagation in a nonlinear diffusion equation.
In this case, we are dealing with a wave equation with
a linear singular perturbation and a specially
ferocious nonlinearity in the stick-slip friction.
Nevertheless, the simplest version of the selection
mechanism seems to be exactly correct for this system:
the selected state is the one for which the
steady-state solution is just on the margin
of being unstable.
While there
remain some technical questions regarding the literal
interpretation of this statement, its practical
implementation has been checked by careful
computations and found to be correct even,  for
example, for large block spacings where the
fourth-derivative approximation would be entirely inaccurate.

	The results of this analysis are indicative of
the interesting structure of the selection problem.
We quote them only in the limit of small $a$ and small
``stuckness'' parameter $\epsilon$: The velocity is
$$
{v\over\xi}\approx 1 + {1\over 2}\left(3\,\alpha\,a\over
2\,\xi\right)^{2/3}\left[1-{\pi^2\over\ln^2(2\epsilon/3)}\right]
\eqno(v)
$$
and the width of the pulse is
$$
\Delta x \approx \sqrt{v^2-\xi^2} \approx \xi\left(3\,\alpha\,a\over
2\,\xi\right)^{1/3}
\eqno(width)
$$
Note the following.  In the continuum limit $a\to 0$,
the velocity approaches the sound speed $\xi$ from
above and thus remains well behaved.  The width
$\Delta x$ vanishes in this limit, but the number of
blocks in it becomes infinite: $\Delta x/a \approx
(\xi/a)^{2/3}\to\infty$. According to \(v), the pulse
slows with increasing $\epsilon$.  This is
qualitatively consistent with the behavior seen in
Fig.6 where there is a clear deceleration as the pulse
passes through regions where the original values of
$U(x)$ are relatively large. At larger values of $\epsilon$, where \(v) and
\(width) are no longer valid, $v$ falls below $\xi$. The pulses fail to
propagate --- the selected velocities vanish --- at $\alpha$-dependent values
of $\epsilon$ that are always somewhat less than unity.

	A specially interesting feature of the results reported in
Myers and Langer [1993a]
is that the selected pulses invariably probe the strongly nonlinear
portion of the friction law \(friction).  That is, the
blocks slip as much as possible at speeds large enough
that the friction is small.

	The idea that the motion in very large earthquakes consists
of narrow propagating slipping pulses has been advanced on
the basis of observational data by Heaton [1990].
Heaton's conjecture
remains a basic open question and is the subject of
considerable debate.
Heaton's picture is supported robustly by our analytic and
numerical studies. Our pulses are narrow in the
sense that their thicknesses $\Delta x$ are appreciably
smaller than the ``crust depth'' $\xi$. Most remarkably,
once they get started, our model pulses move steadily
along the irregularly stressed fault, amplifying irregularities
as they pass, but  with no
noticeable           back-scattering.

\vfill\eject

\head{V. Higher dimensions: moving towards more realistic models}

	Certain intrinsic limitations of the one-dimensional
Burridge--Knopoff model can be overcome by
extending it to higher dimensions.
For example, in the     one-dimensional model,
there are no elastic interactions between
distant points on the fault, and there is no
mechanism for radiative transport of elastic energy.
In addition, real earthquakes have structure
in the two dimensional fault plane and, therefore,
the scaling laws and exponents that we find in one
dimension may not be the same as those in  higher
dimensions.  Of course, the Earth is three dimensional,
but fully three-dimensional calculations analogous to
those we have carried out for the           one-dimensional
model are still beyond the range of our computational
capabilities.  For this reason, our study of higher
dimensional models has begun with the consideration
of two different two-dimensional models representing
two orthogonal cross sections of the seismogenic zone.

	First we consider a two-dimensional fault-plane
model [Carlson, 1991] in which the variable $x$, as before, describes
position along the fault, and $z$ is the depth below the
surface. For simplicity, we consider displacements $U(x,z,t)$
only in the $x$-direction. The equation of motion is
$$
\ddot U =
\xi^2\nabla^2 U-U-\phi(\dot U, z) + \nu\,t, \eqno(2deom)
$$
which differs from Eq.\(eom) in the
 higher dimensional gradient term, representing a
two-dimensional lattice of coupling springs,
and  the depth dependence of the friction.
In the simplest case, we have taken the friction to be
independent of $z$.
A more realistic option is to allow  the friction to
change with depth
to account for the fact that
friction in the Earth depends on pressure and temperature.

	Figure 7 is a typical magnitude vs.~frequency
distribution for the fault-plane model.
Comparing this with the corresponding results for the one-dimensional
model (Fig.5) we see that the scaling law describing small to
moderately large events is remarkably unchanged, that is,
the exponent $b=1$ in the Gutenberg-Richter
relation \(GR) continues to hold. The most striking difference
between the results in one and two dimensions is that,
in two dimensions, the bump associated with large events is
replaced by a relatively flat shoulder. This occurs both
with and without depth dependent friction for systems
that are sufficiently large.  The bump reemerges,
however, for faults that are sufficiently shallow
(roughly for fault depths less than $\tilde\xi$).
In fact, because the Earth's crust is thin in
comparison to the typical propagation length of a great
earthquake, it is likely that the most realistic case
is the relatively shallow two-dimensional
fault with a depth of order $\xi$. In our simulations for
both deep and shallow two-dimensional faults, there
continues to be an excess of large events relative to the
projected rate of smaller events. Of course,
in two-dimensions the crossover $\tilde\mu$ will
be modified because, according to dimensional analysis,
it must scale like $\xi^2$ rather than $\xi$.

	Results from the fault-plane model are
qualitatively comparable to seismic reconstructions [Archuletta, {\it
et al.,} 1982] and
depth dependent measurements [Sibson, 1982]. Figure 8
is an illustration of a moderately large event in
the $x-z$ plane. In Fig.8a we show the blocks that are
slipping at equal time intervals. As in the      one-dimensional model,
we observe narrow propagating fronts.
In this particular case, the event starts near
the bottom of the fault and propagates both horizontally and
vertically, sweeping out a slipping zone that is not at
all radially symmetric or spatially uniform, as illustrated
in the slip distribution shown in Fig.~8b. In fact, in
some cases we observe that the propagating fronts split
and pass around regions that are firmly stuck. In
the shallow-fault model, this splitting tends not
to be observed and, instead, the slipping front
spreads from bottom to top and then
propagates unilaterally or bilaterally along the fault,
again producing an irregular slip distribution.

	Next we consider our second two-dimensional cross
section of the Earth, namely a crustal-plane
model in which a one-dimensional fault (the $x-$axis)
is embedded in a  two-dimensional elastic medium (the $x-y$
plane) [Langer, 1993; Langer and Nakanishi, 1993; Myers and Langer, 1993b].
Again, for simplicity, we consider only a
one-component displacement field $U(x,y,t)$, and we assume
that $U$ satisfies a wave equation with a linear driving force:
$$
\ddot U = \xi^2\nabla^2 U - U + \nu\,t
\eqno(elasteom)
$$
In analogy to \(eom) and \(2deom), the term $(-U+\nu t)$ in
\(elasteom) describes elastic coupling of a seismogenic
layer of thickness $\xi$ to a stable lower region of the crust.
The stick-slip friction between the two sides of the fault is
part of the boundary condition that specifies the
stress $\partial U/\partial y$ at
$y=0$:
$$
\left.\partial U\over \partial y\right|_{y=0}
=\phi(\dot U) - \eta \left.\partial^2 \dot U\over \partial
x^2\right|_{y=0}\eqno(bc)
$$
Note that we have included the viscous damping
mentioned in Sec.IV as the second part of the traction
on the right-hand side of \(bc). This term has the effect
of smoothing the system at the smallest length scales
and thus assuring that  Eqs.\(elasteom) and \(bc)
are well posed mathematically.

	Some recent studies of models of this kind have focussed on the case in which
the stick-slip friction $\phi(\dot U)$ in \(bc) is replaced by a cohesive
stress that depends upon displacement rather than slipping speed. The resulting
model describes ordinary crack propagation with a non-zero fracture energy
rather than unstable rupture on an existing fault. It has some extremely
interesting properties, in particular, a dissipation-dependent threshold for
the onset of rapid motion. The important common feature of both versions of the
model is that they accurately include stress concentration at the crack tip or
rupture front.  This feature is completely absent in the
one-dimensional models, and there is every reason to expect that it should have
a qualitative effect on the dynamics of the system.  Indeed, it is the
combination of            two-dimensional stress concentration and viscous
dissipation on the crack face that produces the interesting properties of the
fra!
cture model.

	So far, studies of this crustal-plane model in earthquake mode
(primarily by C. Myers)
have focussed on issues related to
pulse propagation in analogy to the     one-dimensional studies
described in Sec. IV. Myers has obtained convincing
numerical evidence that large-scale slipping occurs in this
model {\it via} a mechanism ostensibly identical to the
Heaton pulse; the system resticks behind the rupture front,
and the width of the slipping zone is relatively narrow
[Myers and Langer, 1993b].
A picture of one of these pulses is shown in Fig.9.
The major outstanding question is whether this model is
intrinsically chaotic and, if so, whether the earthquake-like
events have magnitudes that are distributed according to
something like the Gutenberg-Richter law.

\head{VI. Predictability:
Forecasting the next large earthquake}

	One of the main goals of seismology is
to develop better algorithms and more sensitive
probes to aid in predicting large earthquakes.
The problem of earthquake prediction is extremely
complex. Our knowledge of the detailed structure
of subsurface fault zones remains remarkably
incomplete, and we have been collecting accurate
information about seismic activity only for a few
decades --- much less than a single loading period
for  most  fault segments. In addition,
earthquakes show  symptoms of being intrinsically
chaotic phenomena.

	The study of dynamical models may
be particularly useful in a situation of this kind.
Models are not limited by geological time scales;
for example, our  computer simulations span the equivalent of
hundreds of loading periods with perfect detection of events. Furthermore,
the uniform Burridge-Knopoff model that we study is
a particularly good candidate for such investigations because
it is deterministically chaotic, making it impossible to
predict detailed behavior far into the future.
Yet this model, realistically, possesses a
characteristic loading time, a distinction between small
and large events, and a tendency ---
quite apparent in Figs.4 and 10 ---
for clusters of small events to be correlated
with the onset of large ones.  Thus it
seems interesting to inquire about the extent to
which it is possible, using only the analogs of
techniques that are available in the real world,
to predict the times and locations of large events in this model.

	The seismological literature refers to three
categories of earthquake hazard assessments: long term,
intermediate term, and short term predictions.  Long term
predictions are estimates of earthquake probability made
roughly on the scale of tens of years [WGCEP, 1988].
Such assessments are used in establishing building
codes and, especially, in siting sensitive
facilities such as nuclear reactors. They generally are
based on estimates of recurrence times  for the large events on major
active faults and on whatever other geological
information might be available.

	In contrast, intermediate term predictions are made
on time scales of years, and short term predictions
on scales of days.  The hope is that some more detailed
information regarding the local state of the system, perhaps based
on patterns of  small  events, may  ultimately be
used to
provide early warnings of imminent large events. Thus there
is much interest in learning how to make intermediate
and short term predictions in a reliable way.
One major contribution in the area of intermediate term prediction
has been made by Keilis--Borok {\it et al.} [1990a, 1990b, 1990c],
who have
developed a set of prediction algorithms  using
relatively simple pattern recognition techniques. However,
because of the sparsity of real seismic data, it has
been difficult to evaluate the quality of these
algorithms.
One purpose of our work has been to estimate the intrinsic uncertainties
and the prospects for  more accurate results
by finding out how well such
algorithms can be made to work for uniform Burridge-Knopoff models [Pepke,
{\it et al.,} 1993].

	The goal of intermediate term earthquake
prediction as formulated by Keilis-Borok {\it et al.}
is considerably more limited than the term might seem to
imply. One is not actually trying to predict earthquakes
ten years in advance.  Rather, the idea is simply to
recognize patterns of seismic activity that might indicate
times of increased probability for major earthquakes --- the
so-called ``TIP's''. The goal is to make these alarms as
accurate as possible.
The times during which these are \lq\lq on" should be short
and the geographical regions to which they apply should be small, and yet
they should \lq\lq capture" almost all of the major seismic events
with few, if any, false alarms.
Given this statement of the
problem, it is relatively easy to define mathematically significant
measures of success and to use those measures to test
various algorithms, that is, to evaluate various criteria
for turning on the TIP's.

In the pattern recognition algorithms of Keilis-Borok {\it et al.}, TIP's are
determined by keeping track of
as many as eighteen different
seismicity-based precursors such as an increase in activity or an
increase in the rate of aftershocks.
The strategy is to use computer-based pattern recognition techniques
to identify various combinations of precursory behavior
that, with high probability, indicate that a major event is imminent.
In the real world,
no single precursor  appears to be
capable of  forecasting all  large events.
The best single precursors can capture roughly half of the
large events but need to turn on alarms at least 20\% of the
recurrence interval in order to do so.
When combinations of seven or more precursors are used, the success
rate goes to about 80\%, but with the same, relatively poor, time
resolution.

The patterns of activity that precede large events in the
uniform Burridge-Knopoff
model are much simpler than in the real
world; thus  we expect the
pattern recognition algorithms to be
comparatively more successful.
In particular, almost every large event in the model is
preceded by a marked increase in activity, i.e.~the rate of small
events,
in the neighborhood of the future epicenter
[Shaw {\it et al.,} 1992].
This is illustrated in Fig.~10 which, like Fig.~4, illustrates
a typical sequence of events.
In this case, the time of occurrence of each event
appears  explicitly, and  each event is marked by a line
drawn through the blocks that slip. For large events, a
cross marks the
position of the epicenter, which is clearly correlated
with precursory activity. However, the duration of
this activity is relatively long-- about one third of the entire repeat
time on average. Moreover, there is large
variability in the overall amplitude of the precursory activity, and
in the time prior to the large event at which it starts.
Thus the question remains: to what extent is this precursory activity
useful for making predictions on the times scales  (a small fraction
of the mean repeat time) that are
useful for intermediate term forecasts?

To  answer this question,
we have considered the simplest versions of the
pattern recognition algorithms, that is, algorithms based on  single
precursors.
We find that with most precursors such as  the total activity, we can
predict roughly 90\% of the
events with alarm times of order 15\%--20\% of the seismic
cycle.
This is somewhat better than  Keilis--Borok {\it et al.}'s
real-world experience,
but not dramatically so.
Interestingly, there is one exception to this. We have found
a  new precursor, not previously
used by seismologists, that
leads to significantly more accurate
predictions in the Burridge--Knopoff
model. The definition of this  new precursor, which we call
\lq\lq active
zone size," is related to the concept of the delocalization length
$\tilde\xi$. With it we can predict 90\% of the large events
with TIP's that are \lq\lq on" only during
5\% of the recurrence interval.

\lq\lq Active zone size"
is the spatial extent of small-scale seismicity.
In a space-time window  such as that illustrated in Fig.~10, the
active zone size
is  defined to be the total number of blocks that have
slipped, regardless of how many times. As the time of a large event
approaches, the active zone grows,  leading to the development
of a triggering region of size  $\tilde\xi$
for the nucleation of a
large event,  as  described in Section IV.
Since nearly all of the accumulated stresses are relieved
in the large events,
the small  events serve primarily as indicators
that the system is locally poised near the threshold of instability.
While
the total number of events (i.e. activity) is  clearly not an  independent
measure of the status of the system,
it is a much less sensitive  probe of the
size of the region that is close to threshold.
It remains to be seen
whether or not this new measure will aid in efforts to
predict large events in the earth.

Many  questions pertaining to the use of
models in  the development and
testing of  earthquake prediction
algorithms are still unanswered.
One of these is the extent to which algorithms may be
improved by combining multiple precursors as is done by
Keilis-Borok and colleagues.
We  currently are
studying this question and, so far, find that such techniques do not
lead to
substantial gains in the Burridge--Knopoff model.
In fact, it
is not immediately clear to what extent even comparatively sophisticated
techniques might be successful if based only on information
contained in our catalogs of seismic events.
More accurate predictions may require information about
other phenomena such as aftershocks,
which do not occur in
our current version of the model,
but which  we hope to add  in the future
[Shaw, 1993b]. Because the model is fully deterministic,
arbitrarily accurate
predictions would be possible
if more detailed information about the  configuration of the system
could somehow be obtained.
For example, we would like to test the possibilities
of combining seismicity data with the analog in the model of geological
information about local displacements or stresses. It is this kind of
investigation that we suspect may lead to the
most significant practical applications of the work done in
this project.

\noindent
$Acknowledgements$:
We would especially like to thank our collaborators  who made
important contributions to the work presented here:
C. Tang, C. Myers,  S. Pepke, and H. Nakanishi.
The work of JMC was supported by a grant from the Alfred P.
Sloan Foundation,  NSF grant DMR-9212396, and
an INCOR grant from the CNLS at
Los Alamos National Laboratories.
The work of JSL was supported by  DOE grant DE-FG03-84ER45108.
The work of BES was supported by the SCEC grant USC-572726,
and USGS grant 1434-93-G-2284.
The work of JMC, JSL and BES  was also supported  by
NSF grant
PHY89-04035.

\vfill\eject

\centerline{REFERENCES}

\noindent
Archuletta, R.J., E. Cranswick, C. Mueller, and P. Spudich,
1982,
{\it J. Geophys. Res.} {\bf 87},  4595.

\noindent
Bak, P., C. Tang, and K.
Wiesenfeld,  1987,
{\it Phys. Rev. Lett.} {\bf 59}, 381.

\noindent
Barriere, B., and D.L. Turcotte, 1991,
{\it Geophys. Res. Lett.} {\bf 18},
2011.

\noindent
Burridge, R., and L. Knopoff, 1967,
{\it Bull.
Seismol. Soc. Am.} {\bf 57}, 3411.

\noindent
Carlson, J.M.,  1991,
{\it J. Geophys. Res.} {\bf  96}, 4255.

\noindent
Carlson, J.M., 1991,
{\it Phys. Rev. A} {\bf 44},
6226.

\noindent
Carlson, J.M., J.S. Langer, B.E. Shaw, and C. Tang, 1991,
{\it Phys. Rev. A} {\bf 44},
884.

\noindent
Carlson, J.M. and J.S. Langer, 1989a,
{\it Phys. Rev. Lett.} {\bf 62}, 2632.

\noindent
Carlson, J.M. and J.S. Langer, 1989b,
{\it Phys. Rev. A} {\bf 40}, 6470.

\noindent
Chen, K., P. Bak, and S. Obukov, 1991,
{\it Phys. Rev. A} {\bf 43}, 625.

\noindent
Davison, F.C., Jr.,
and C.H. Scholz, 1985,
{\it Bull. Seismol. Soc. Am.} {\bf 75}, 1349.

\noindent
Glatzmaier, G.A., G. Schubert, and D. Bercovici,
1990,
{\it Nature} {\bf 347}, 274.

\noindent
Gutenberg, B., and C. F. Richter, 1954,
{\it Seismicity of the Earth and Associated Phenomena,}
Princeton University Press, Princeton, NJ, 310pp.

\noindent
Heaton, T.H., 1990,
{\it Phys. Earth Plan Inter.} {\bf 64}, 1.

\noindent
Kagan, Y.Y., 1982,
{\it Geophys. J. R. Astr. Soc. 71}, 659.

\noindent
Keilis-Borok, V.I., and I.M. Rotwain,  1990a,
{\it Phys. Earth Planet. Inter.} {\bf 61}, 57.

\noindent
Keilis-Borok, V.I., and V.G. Kossobokov,  1990b,
{\it Phys. Earth Planet. Inter.} {\bf 61},
73.

\noindent
Keilis-Borok, V.I., L. Knopoff, V.G. Kossobokov, and I. Rotwain, 1990c,
{\it Geophys. Res.
Lett.} {\bf 17}, 1461.

\noindent
Knopoff, L., 1993, unpublished.

\noindent
Kolmogorov, A., I. Petrovsky, and N. Piscounov, 1937,
{\it Bull. Univ. Moscow Ser. Internat. Sec. A} {\bf 1}, 1.

\noindent
Langer, J.S. and C. Tang, 1991,
{\it Phys. Rev. Lett.} {\bf 67}, 1043.

\noindent
Langer, J.S., 1992,
{\it Phys. Rev. A} {\bf  46}, 3123.

\noindent
Langer, J.S., 1993,
{\it Phys. Rev. Lett.} {\bf  70}, 3592.

\noindent
Langer, J.S., and H. Nakanishi, 1993,
{\it Phys. Rev. E}, to appear.

\noindent
Myers, C.R.,
and J.S. Langer,  1993a,
{\it Phys. Rev. E} {\bf 47}, 3048.

\noindent
Myers, C.R. and  J.S. Langer, 1993b, in preparation.

\noindent
Nishenko, S.P., and R. Buland, 1987,
{\it Bull. Seismol. Soc. Am., 77}, 1382.

\noindent
Pacheco, J.F., C.H. Scholz, and L.R. Sykes, 1992,
{\it Nature, 235}, 71.

\noindent
Pepke, S.L., J.M. Carlson, and B.E. Shaw, 1993,
ITP preprint number NSF-ITP-93-71.

\noindent
Sahimi, M., M.C. Robertson, and C.G. Sammis, 1993,
Relation between earthquake statistics and fault patterns, and fractals
and percolation, (unpublished).

\noindent
Scholz, C.H., 1990, {\it The mechanics of earthquakes and faulting},
Cambridge University Press, New York, 435pp.

\noindent
Schwartz, D.P. and K.J. Coppersmith, 1984,
{\it J.
Geophys. Res.} {\bf 89}, 5681.

\noindent
Shaw, B.E., J.M. Carlson, and J.S. Langer, 1992,
{\it J. Geophys. Res.} {\bf 97}, 479.

\noindent
Shaw, B.E., 1993a,
{\it Geophys. Res. Lett.} {\bf 20}, 643.

\noindent
Shaw, B.E.,  1993b,
{\it Geophys. Res. Lett.} {\bf 20}, 907.

\noindent
Sibson, R.H.,  1982,
{\it Bull. Seis. Soc. Amer.} {\bf 72},
151.

\noindent
Vasconcelos, G.L., M.S. Vieira, and S.R. Nagel, 1992,
{\it Physica A} {\bf 191}, 69.

\noindent
Wesnousky, S.G., 1993,
\lq\lq The Gutenberg-Richter or characteristic
earthquake distribution, which is it?," (unpublished).

\noindent
Working Group on California Earthquake Prediction (WGCEP), 1988,
{\it U.S. Geol.
Sur.  Open File Rep. } {\bf 88-398}.

\endreferences
\vfill\eject

\vfill\eject

\centerline{FIGURES}

\noindent
(1). A schematic representation of a fault.
Shear and normal stresses are applied far from the interface
between the homogeneous elastic plates, and are relieved by sudden
slipping motions which occur when the friction threshold is
exceeded.

\noindent
(2). The one-dimensional uniform Burridge--Knopoff model represents the
finite difference approximation to Eq.~(1) and
consists of a chain of elastically coupled blocks,  each of which is
also coupled to a slowly moving surface.
The
key nonlinearity leading to complex behavior
is the friction
law illustrated in Fig.~3.

\noindent
(3). The  velocity-weakening slip-stick friction law.

\noindent
(4). For a typical sequence of events,
we plot the displacement $U(x,t)$
immediately after each event,
beginning after the initial transient
has passed, so that the system has reached a statistically steady state.
In spite of the underlying homogeneity of the
system, the behavior is quite complex. While the system is
technically chaotic, it also displays
short term \lq\lq patterns"  which can be used for predictive purposes.
Here we have taken $\sigma=.01$, $\alpha=1.2$, $\xi/a=2.$ and $N=500$.
In the more generic case of $\alpha> 2$ the small events account
for even less of the net displacement than for the parameter values taken here.

\noindent
(5). Magnitude vs.~frequency distribution for the uniform
Burridge--Knopoff model.  Here we have taken
$\sigma=.01$, $\alpha=2.5$, $\xi/a=6$ and $N=1500$.
A more detailed numerical study of the scaling of this distribution
reveals an interesting sensitivity of  the magnitude of the
largest events to presence of the  short-wavelength cutoff
$a$
[Carlson
{\it et al.,} 1991].
In particular, for $N$ sufficiently large,
both the magnitude of the  crossover  $\tilde\mu$  and
largest event $\mu^*$  are independent of $N$. However,
while $\tilde \mu$ is independent of $a$, we find that
$\mu^*$  increases as the mesh becomes finer, scaling roughly
as
$\mu^* \sim\ln( \xi^2/a)$.

\noindent
(6). The dynamics of an individual large event consist of
narrow slip pulses which propagate at  a speed of roughly $\xi$.
New curves are drawn at equal time intervals during the event,
and represent the configuration at various stages between the initial
configuration (bottom curve) and final configuration (top curve).
The spatial variation of the pulse speed is apparent in the
variation in spacing of the nearly vertical lines which mark the
slipping blocks. Lines which are relatively farther apart
correspond to fast moving pulses which propagate through regions which are
close to threshold, while the lines that are closer indicate a
relatively slower speed  which occur
when the pulses pass through regions which
are more stuck.
It is interesting to note that as a consequence of the large event, the
configuration is nearly inverted:  regions which were
initially close to threshold slip further than regions which were
initially  far  from threshold.
Here we have taken $\xi/a=3$, $\alpha=1.2$ and $\sigma=.01$.

\noindent
(7). Magnitude vs.~frequency distribution in the two-dimensional
fault plane model. Here we have taken $\sigma=.01$, $\alpha=2.5$,
$\xi/a=3$, $N_x=200$ and $N_z=100$.
The results shown here are for depth dependent friction. The
corresponding distribution
obtained when friction does not depend on depth is essentially
identical.

\noindent
(8). A typical  delocalized event in the two-dimensional fault-plane model
(parameters  as in Fig.~7). This event has $\mu=6$,
which on the corresponding magnitude vs.~frequency distribution lies
a little more than
half of the way
between $\tilde\mu$ and the largest event  $\mu^*$.
In this event 3522 of the 20,000 blocks were involved. The top
figure illustrates the slip distribution during the particular event.
Black corresponds to no slip, while the
grey to white scale ranges linearly up to the maximum displacement
of $0.5$ during the event.
The bottom figure illustrates the slipping blocks at equal intervals
of time ($\Delta t=5$, and the total duration is $t=44$), grey
corresponding to early times and white corresponding to the latest
times during the event.

\noindent
(9).
Snapshot of the slip rate $\dot U(x+vt,y)$ in the two-dimensional crustal
plane model, demonstrating the existence of a ``self-healing pulse
of slip".  The fault lies along the line $y=0$, and the snapshot is
taken in a co-moving frame of velocity $-v$, where $v=1$ is the
steady-state velocity for this particular pulse.  The pulse is propagating
to the right along the $x+vt$ axis.  The model parameters are
$\alpha = 3.0, \eta = 0.1$. A first-order implicit finite-difference
integration is carried out on a rectangular grid of $250\times100$ points
(with $\Delta x = 0.01, \Delta y = 0.05$), although only 80 points are shown
in the $y$-direction.  The boundaries are free (zero derivative boundary
conditions) along all edges except $y=0$, where the fault traction is
specified by the model.  An artificial viscous zone is introduced for
$y > 4.5$ in order to minimize reflections off the back boundary.
In this zone, an extra viscous term of the form $\eta(y) \nabla^2(\dot U)$
is added to the bulk equation of motion, with $\eta(y)$ growing smoothly
from zero.

\noindent
(10). A sample catalog illustrating the events which take place as a
function of space and time.
Time on the vertical axis is measured  relative to  the inverse loading
speed,
so that values on that axis represent  $t\nu=\delta U$, i.e.~the
net  displacement of initially adjacent points
on opposite sides of the fault.
A line segment  is drawn through all of the
blocks which slip during an event, and a cross marks the position of the
epicenter of each large event.
The data is taken from a simulation in which
$\sigma=.01$, $\alpha=3$, $\xi/a=10$, and
$N=8192$,
though only a  fraction of the system is shown.

\end